\documentclass[preprint,showkeys,aps]{revtex4-1}
\usepackage{epsf}
\usepackage{amsmath,amsthm,amssymb}
\usepackage{color}
\usepackage{dcolumn}
\usepackage{bm}
\usepackage{graphicx,epsfig}
\usepackage[dvipsnames]{xcolor}

\graphicspath{{fig/}}
\everymath{\displaystyle}

\begin{document} 

\title{Effective oscillatory magnetic charges and electric dipole moments induced by axion-photon coupling}

\author{Alexander J. Silenko$^{1,2}$}
\email{alsilenko@mail.ru}

\affiliation{$^1$Bogoliubov Laboratory of Theoretical Physics, Joint
Institute for Nuclear Research, Dubna 141980, Russia}
\affiliation{$^2$Moscow Center for Advanced Studies, 
Kulakova str. 20, Moscow 123592, Russia}

\date{\today}

\begin{abstract}
We suppose that the covariance of Lagrangians always taking place results in a distortion of any electric and magnetic field by a pseudoscalar field of dark matter axions and axion-like particles. As a result, electric and magnetic fields acquire oscillating magnetic and electric components, respectively. The Maxwell-like equations are rigorously derived. One can also use an equivalent approach based on an introduction of effective oscillating magnetic charges and electric dipole moments in undistorted electromagnetic fields. The determined relations between the magnetic charges and electric dipole moments and the axion-photon coupling constant open a possibility to compare a sensitivity of search for axions in optical experiments and experiments with massive particles. The dual (inverse) Witten effect defining effective magnetic charges of electrically charged particles opens new exciting possibilities to search for axions and axion-like particles with very small masses in nonresonance experiments. The study of a passage of strongly decelerated electrons or positrons through a solenoid is proposed.
\end{abstract}


\maketitle

The axion is a hypothetical neutral pseudoscalar particle with a very low mass $m_a\lesssim10^{-2}$eV/$c^2$ postulated by Peccei and Quinn \cite{PecceiQuinn1,PecceiQuinn2}. It is a quantum of the pseudoscalar field. Its existence resolves the strong \emph{CP} problem in quantum chromodynamics (QCD). If the axion or an axion-like particle (ALP) exists, it can be a possible component of cold dark matter. The mass of dark matter axions is restricted by cosmological arguments \cite{amasstheor}, astrophysical observations \cite{amassgamma,amassobs}, and quantum field theory \cite{Weinberg,DiLuzio,AxionMassEstimates,AxionMassNatureComm}. An experimental search for axions and ALPs can result in a discovery of fifth force. 

One of the most important manifestations of the axion (being the pseudoscalar field quantum) is an appearance of time-dependent electric dipole moments (EDMs) of nuclei and strongly interacting particles \cite{GrahamPRD2,GrahamPRDgen,Grahamannurev}. Such EDMs appear due to the axion-gluon coupling and have been appropriately studied (see Refs. \cite{Grahamannurev,PhysUsp,axiondJEDI} and references therein). However, these investigations do not relate to 
axion-induced EDMs of leptons. The successful attempt of theoretical determination of such EDMs based on the analysis of a Feynman diagram has been undertaken in Ref. \cite{Hill}. It has been noted that the axion-induced electron EDM is proportional to the Bohr (Dirac) magnetic moment $\mu_0=e/(2m)$. The result \cite{Hill} has been debated in Refs. \cite{FlambaumComment,AxionHillReply,AxionHill2016}.

Witten \cite{EWitten} has obtained that a \emph{CP}-violating vacuum angle $\theta$ leads to an acquisition of electric charges by 't Hooft-Polyakov monopoles:
\begin{equation}
\begin{array}{c}
q=-\frac{e\theta}{2\pi}+ne,
\end{array}
\label{ewitten}
\end{equation} where $n$ is an integer. The wonderful research \cite{EWitten} has been used and developed in plenty of publications. We can point out Refs. \cite{DiLuzio,Fischler,Niemi} and recent papers where EDMs caused by the Witten effect have been stated \cite{AxionCao,AxionZhitnitsky} and quantum electromagnetodynamics 
has been presented \cite{AxionLi,AxionLiArxiv,AxionLiZhang,SokolovPRD,SokolovAxion,SokolovRingwald,Heidenreich}.

Some different effects caused by axion-photon interaction have been claimed in Refs. \cite{Alexander,Wang}.

We use the standard denotations of Dirac matrices (see, e.g., Ref. \cite{BLP}) and the Gaussian units with $\hbar=1,~c=1$. In these units, the absolute dielectric permittivity and the magnetic permeability of vacuum are equal to $1/(4\pi)$ and
$4\pi/c^2=4\pi$, respectively. We include $\hbar$ and $c$ explicitly when this inclusion
clarifies the problem.

We consider mixing electric and magnetic fields in a pseudoscalar axion
background field, which is tantamount to endowing particles with 
oscillating magnetic charges and EDMs proportional to the axion field.
We confirm the proportionality of the axion-induced EDM to the magnetic dipole moment \cite{Hill,AxionHillReply,AxionHill2016,AxionCao,AxionZhitnitsky} and the existence of the Witten effect. However, our approach draws a very different physical picture. We obtain that axion-induced EDMs, electric charges of monopoles, and magnetic charges of electrically charged particles (dual or inverse Witten effect) are effective and even fictitious \cite{footnote}. 
In accordance with the approach used in all previous studies, electromagnetic fields and their potentials are not distorted by axions and ALPs. In contrast, our approach predicts mixing the electric and magnetic fields and the change of field potentials. We take into account only axion-photon coupling $g_{a\gamma\gamma}$ distorting the Lagrangian density of a free electromagnetic field. The constant $g_{a\gamma\gamma}$ is model-dependent (see Ref. \cite{Grahamannurev} for more details). The distorted Lagrangian density is given by
\begin{equation}
\begin{array}{c}
{\cal L}'=-\frac{1}{16\pi}{F'}_{\mu\nu}{F'}^{\mu\nu}={\cal L}+{\cal L}_{\gamma},\qquad {\cal L}=
-\frac{1}{16\pi} F_{\mu\nu}F^{\mu\nu},\\{\cal L}_{\gamma}=-\frac{g_{a\gamma\gamma}}{16\pi}{a(x)}F_{\mu\nu}\widetilde{F}^{\mu\nu}=\frac{g_{a\gamma\gamma}}{4\pi}a(x)\bm E\cdot\bm B,\qquad \widetilde{F}_{\mu\nu}=\frac12\epsilon^{\mu\nu\alpha\beta}F_{\alpha\beta},
\end{array}
\label{eq33nnw}
\end{equation} where ${\cal L}$, $F_{\mu\nu}=\partial_\mu A_\nu-\partial_\nu A_\mu$, and $A_\mu$ are the Lagrangian density, the electromagnetic field tensor, and the four-potential without the axion field. The tilde denotes a dual tensor, $\epsilon^{\mu\nu\alpha\beta}$ is the Levi-Civita symbol, and $x$  stands for four spatial and temporal coordinates. We neglect terms of the second order in $a(x)$. Like photons, moving axions form a wave which pseudoscalar field is defined by \cite{GrahamPRDgen2018}
\begin{equation} a(x)\equiv a(\bm r,t)=a_0\cos{(E_at-\bm p_a\cdot\bm r+\phi_a)}, \label{axion}
\end{equation} where $E_a=\sqrt{m_a^2+\bm p_a^2}$, $\bm p_a$, and $m_a$ are the energy, momentum, and mass of axions. Since $a(x)$ depends on four coordinates, ${\cal L}_{\gamma}$ cannot be a total derivative except for the trivial case of $m_a=0,~\bm p_a=0$. 
As a result, the Lagrangian density ${\cal L}_{\gamma}$ is not unphysical. 
\emph{CP}-noninvariant interactions caused by dark matter axions are time-dependent. The Earth motion through our galactic defines its velocity relative to dark matter, $V \sim10^{-3}c$. Therefore, $|\bm p_a|\approx m_a V$ \cite{GrahamPRDgen} and axions and ALPs have momenta of the order of $|\nabla a(\bm r,t)|\sim 10^{-3}\dot{a}(\bm r,t)c$. Since  $|\bm p_a|\ll m_ac$, one mostly uses the approximate equation
\begin{equation} a(x)=a_0\cos{(m_at-\bm p_a\cdot\bm r+\phi_a)}. \label{axionap}
\end{equation}
\textcolor{magenta}{
}

In some papers (see Refs. \cite{Weinberg,DiLuzio,AxionMassEstimates,AxionMassNatureComm} and references therein), the axion mass has been evaluated. It has been determined in the review \cite{DiLuzio} on the base of Ref. \cite{Weinberg} that
\begin{equation}
m_{a}\simeq 5.7\left(\frac{10^{12}\, \rm{GeV}}{f_a}\right)\mu\rm{eV},
\label{eqWeL}
\end{equation} where $f_a$ is the axion decay constant.
The estimate $m_a=(110\pm2)$ $\mu$eV has been found in Ref. \cite{AxionMassEstimates}. The restriction $(40\geq m_a\leq180)$ $\mu$eV has been obtained in Ref. \cite{AxionMassNatureComm}. The axion momentum is rather small and corresponds to the wavelength $\lambda_a=h/|\bm p_a|\sim10$ m. Such a wavelength is much larger than sizes of a most part of laboratory devices. 
Therefore, hereinafter we neglect terms with derivatives $\nabla a(\bm r,t),~\dot{a}(\bm r,t)$ in equations of motion for the particle momentum and spin and terms with spatial derivatives $\nabla a(\bm r,t)$ in Maxwell-like equations. Due to the smallness of the axion mass and momentum, such terms are small as compared with terms taking into account.

We should distinguish between $\lambda_a$ which is the de Broglie wavelength and the Compton wavelength $\lambda^{(C)}_a=h/(m_ac)$. Only the wavelength $\lambda_a$ defines interference properties of axion waves. We underline that Eq. (\ref{axionap}), unlike Eq. (\ref{axion}), is inapplicable in the limit of a vanishing axion mass ($m_a\rightarrow0$). This is important in view of the fact that massive pseudoscalar and scalar fields should smoothly pass into the corresponding massless ones in the limit $m_a\rightarrow0$. The use of approximate Eq. (\ref{axionap}) can create an illusion that a transition to this limit eliminates all axion-induced effects. However, Eq. (\ref{axion}) 
shows that a smooth transition to the case of $m_a=0$ takes place. In this case, a massless particle has a nonzero total energy $E_a=cp_a>0$ and can exist. While the axion mass is nonzero, it can be rather low (see, e.g., Ref. \cite{AxionOuellet}) and the availability of such a transition can be significant.

The interesting case of vanishing the mass of \emph{nonrelativistic} axions [$m_a\rightarrow0,~p_a/(m_ac)\ll1$] needs a separate consideration. In this limiting case, ${\cal L}_{\gamma}$ is a total derivative. It is however evident that this case is characterized by $m_a=0, p_a=0$, and $E_a=0$. Therefore, one passes to a massless particle with a zero momentum and zero total energy. Such a particle is unobservable and nonexistent. As a result, a ``smooth'' passage to the limit $m_a \to 0$ at fixed $p_a/(m_ac)\sim10^{-3}$ is that from the existent to the nonexistent particle and can be disregarded.

The distortion of the electromagnetic field tensor equivalent to that 
of the Lagrangian density is given by
\begin{equation}
\begin{array}{c}
F'_{\mu\nu}=F_{\mu\nu}+\frac{g_{a\gamma\gamma}}{2}a(x)\widetilde{F}_{\mu\nu}.
\end{array}
\label{eqtrnsf}
\end{equation}
The derivation of Eq. (\ref{eqtrnsf}) is perfectly rigorous because it is based only on an applicability of the Lagrangian approach to the axion field.

Since $F_{\mu\nu}=(\bm E,\bm B)$ and $\widetilde{F}_{\mu\nu}=(\bm B,-\bm E)$, \emph{any} electric and magnetic fields are transformed as follows: 
\begin{equation}
\begin{array}{c}
\bm E'=\bm E+\frac{g_{a\gamma\gamma}}{2}a(x)\bm B,\qquad \bm B'=\bm B-\frac{g_{a\gamma\gamma}}{2}a(x)\bm E.
\end{array}
\label{eqtrnfl}
\end{equation} 
$\bm E'$ and $\bm B'$ are real electromagnetic fields at the presence of dark matter axions. 

It has been accepted in \emph{all} precedent studies (see Refs. \cite{Sikivie,AxionLi,AxionLiArxiv,AxionLiZhang,SokolovAxion,SokolovRingwald,Heidenreich,AxionOuellet} and references therein) that there exist the usual electromagnetic field and the additional field defined by the Lagrangian densities ${\cal L}$ and ${\cal L}_{\gamma}$, respectively. In this approach, electromagnetic fields and their potentials are not distorted by axions but ${\cal L}_{\gamma}$ affects electrodynamics. Our approach predicts that axions turn the electric and magnetic fields into the mixed fields $\bm E'$ and $\bm B'$, respectively. The axion-photon coupling does not change electromagnetic field sources (charges, currents, and multipole moments). We underline the covariance of the electromagnetic field tensor and the Lagrangian densities ${\cal L}$ and ${\cal L}'$ substantiating the applicability of these quantities to \emph{any} electromagnetic field. In this physical picture, Eqs. (\ref{eqtrnsf}) and (\ref{eqtrnfl}) define also the four-potential $A'_\mu=(A'_0,-\bm A')$ because ${F'}_{\mu\nu}=\partial_\mu A'_\nu-\partial_\nu A'_\mu$. It can be seen that $\bm E'=-\nabla A'_0-\partial_0 \bm A',~\bm B'=\nabla\times\bm A'$. The four-potential $A'_\mu$ should correspond to the electromagnetic field tensor ${F'}_{\mu\nu}$ and should be introduced into the Lagrangian and Hamiltonian instead of $A_\mu$. This is a key difference between our approach and the traditional one.

The Maxwell equations can be unambiguously derived from the \emph{total} Lagrangian density. An applicability of these equations for not only the light field but also electric and magnetic fields of other origins underlines the covariance of Lagrangians. If Lagrangians were not covariant, the Lagrangian method would be inapplicable. As a result, fields of laboratory devices such as condensers and solenoids, and also electromagnetic fields emitted by oscillating charges, currents, and dipoles are transformed by the axion field. The condenser and solenoid fields contain oscillating magnetic and electric components, respectively.

$A'_\mu$ does not explicitly enter into equations of motion for the kinetic momentum and the spin due to the gauge freedom. The kinetic momentum takes the form $\bm\pi=\bm p-e\bm A'$, where $\bm p$ is the canonical (generalized) momentum. 

The connection between the four-potential and external fields has the form 
\begin{equation}
\begin{array}{c}
-\nabla A'_0-\partial_0 \bm A'=\bm E'=\bm E+\frac{g_{a\gamma\gamma}}{2}a(x)\bm B,\\ \nabla\times\bm A'=\bm B'=\bm B-\frac{g_{a\gamma\gamma}}{2}a(x)\bm E.
\end{array}
\label{eqnptnt}
\end{equation}

As follows from the definitions of fields (\ref{eqnptnt}), the calculation of $\nabla\times\bm E'$ and $\nabla\cdot\bm B'$ results in the uniform Maxwell-like equations \begin{equation}
\nabla\times\bm E'+\frac{\partial\bm B'}{\partial t}=0,\qquad \nabla\cdot\bm B'=0.
\label{eqMlEBm} \end{equation} The uniform Maxwell-like equation (Bianchi identity) can be deduced (see Ref. \cite{Jackson}) from these equations and is given by
\begin{equation}
\partial_\lambda F'_{\mu\nu}+\partial_\mu F'_{\nu\lambda}+\partial_\nu F'_{\lambda\mu}=0.
\label{eMo}
\end{equation} 

In the traditional approach, the calculation of $\nabla\times\bm E$ and $\nabla\cdot\bm B$ in the equations defining the electromagnetic fields,
\begin{equation}\bm E=-\nabla A_0-\partial_0 \bm A,\qquad\bm B=\nabla\times\bm A,\label{potfi}
\end{equation}
results in the well-known uniform Maxwell equations:
\begin{equation}
\nabla\times\bm E+\frac{\partial\bm B}{\partial t}=0,\qquad \nabla\cdot\bm B=0.
\label{Mwe}
\end{equation}

In the four-dimensional form, the uniform Maxwell equation (Bianchi identity) reads \cite{Jackson}
\begin{equation}
\partial_\lambda F_{\mu\nu}+\partial_\mu F_{\nu\lambda}+\partial_\nu F
_{\lambda\mu}=0.
\label{uMe}
\end{equation}

Equations (\ref{eqnptnt}) -- (\ref{uMe}) should be commented on. The form of Eqs. (\ref{potfi}) -- (\ref{uMe}) is defined only by the connection between the fields and potentials. Since this connection is independent of a chosen approach, Eqs. (\ref{potfi}) -- (\ref{uMe}) are valid not only for the traditional approach but also for our one.

The derivation of other Maxwell and Maxwell-like equations is straightforward. One should 
use the total Lagrangian density of electromagnetic field which includes interactions with external field sources:
\begin{equation}
{\cal L}_{tot}={\cal L}-A_\mu j^\mu,\qquad j^\mu=(\rho,\bm j),
\label{eqtenlf}
\end{equation} where $j^\mu$ is the current density four-vector and $\rho$ is the charge density. \textcolor{magenta}{In our approach,} the corresponding distorted Lagrangian density reads
\begin{equation}
{\cal L}'_{tot}={\cal L}'-A'_\mu j^\mu.
\label{eqadist}
\end{equation}  
The use of the Euler-Lagrange equation allows one to deduce the Maxwell equations from the Lagrangian density (\ref{eqtenlf}). The same approach applied to the Lagrangian density (\ref{eqadist}) leads to the Maxwell-like equation:
\begin{equation}
\partial_\mu F'^{\mu\nu}=4\pi j^\nu.
\label{eMl}
\end{equation}  

Two Maxwell-like equations in vector form read
\begin{equation}
\nabla\cdot\bm E'=4\pi\rho,\qquad \nabla\times\bm B'=4\pi\bm j+\frac{\partial\bm E'}{\partial t}.
\label{eqMlEBp}
\end{equation} 
As follows from Eqs. (\ref{eqtrnfl}), (\ref{eqMlEBm}), and (\ref{eqMlEBp}),
\begin{equation}
\nabla\cdot\bm E=4\pi\rho,\qquad \nabla\cdot\bm B=2\pi g_{a\gamma\gamma}a(x)\rho.
\label{eqMlEBw}
\end{equation} 
Similarly to Eq. (\ref{eqMlEBw}), we obtain
\begin{equation} \begin{array}{c}
\nabla\times\bm E=-\frac{\partial\bm B}{\partial t}-2\pi g_{a\gamma\gamma}a(x)\bm j+\frac{g_{a\gamma\gamma}}{2} \dot{a}(x)\bm E,\\ \nabla\times\bm B=4\pi\bm j+\frac{\partial\bm E}{\partial t}+\frac{g_{a\gamma\gamma}}{2} \dot{a}(x)\bm B.
\end{array} \label{eqMEB}
\end{equation}
Equations (\ref{eqMlEBw}) and (\ref{eqMEB}) perfectly agree with Eqs. (\ref{eqtrnsf}) and (\ref{eqtrnfl}).

The Euler-Lagrange equation was used for a derivation of Maxwell-like equations in all papers devoted to axion electro(magneto)dynamics. Maxwell-like equations in the presence of the axion field have been first obtained by Sikivie \cite{Sikivie}. The approach used in Ref. \cite{Sikivie} and \emph{all} other precedent studies (see Refs. \cite{AxionLi,AxionLiArxiv,AxionLiZhang,SokolovAxion,SokolovRingwald,Heidenreich,AxionOuellet} and references therein) is based on the four-current coupling with the undistorted four-potential, $-A_\mu j^\mu$. In these cases, the Maxwell-like equations are explicitly \cite{AxionOuellet} or implicitly deduced from the Lagrangian density 
\begin{equation}
{\cal L}''_{tot}={\cal L}+{\cal L}_{\gamma}-A_\mu j^\mu.
\label{eqold}
\end{equation}
In some papers \cite{SokolovPRD,AxionLi,AxionLiArxiv,AxionLiZhang,SokolovAxion,SokolovRingwald}, additional axion-photon interactions were introduced. In our approach, fields created by any source are distorted by the axion field in accordance with Eqs. (\ref{eqtrnsf}) and (\ref{eqtrnfl}). 
In the discussion on magnetic charges and the dual Witten effect in axion electrodynamics \cite{SokolovPRD,AxionLi,AxionLiArxiv,AxionLiZhang,SokolovAxion,SokolovRingwald,Heidenreich}, only the approach based on the Lagrangian density $-A_\mu j^\mu$ (but not $-A'_\mu j^\mu$) has been taken into consideration. It has been concluded in the recent publication \cite{Heidenreich} on the basis of this approach that non-standard axion electrodynamics including the dual Witten effect is not phenomenologically viable. However, this conclusion does not relate to our approach.

The use of the Euler-Lagrange equation for the Lagrangian density (\ref{eqold}) leads to the following Maxwell-like equation:
\begin{equation}
\partial_\mu \left[F^{\mu\nu}+g_{a\gamma\gamma}a(x)\widetilde{F}^{\mu\nu}\right]=4\pi j^\nu.
\label{cMleq}
\end{equation}
The corresponding vector equations read
\begin{equation}
\nabla\cdot\left[\bm E+g_{a\gamma\gamma}a(x)\bm B\right]=4\pi\rho,\qquad \nabla\times\left[\bm B-g_{a\gamma\gamma}a(x)\bm E\right]=4\pi\bm j+\frac{\partial}{\partial t}\left[\bm E+g_{a\gamma\gamma}a(x)\bm B\right].
\label{eqMltra}
\end{equation}
While Eqs. (\ref{cMleq}) and (\ref{eqMltra}) seem to be similar to our Eqs. (\ref{eMl}) and (\ref{eqMlEBp}), they are coupled to Eqs. (\ref{Mwe}) and (\ref{uMe}). As a result, one comes to the well-known equations of axion electrodynamics \cite{Sikivie,AxionLi,AxionLiArxiv,AxionLiZhang,SokolovAxion,SokolovRingwald,Heidenreich,AxionOuellet}. In the considered case, the spatial derivatives of $a(x)$ are neglected and these Maxwell-like equations have the form 
\begin{equation} \begin{array}{c}
\nabla\cdot\bm E=4\pi\rho,\qquad \nabla\cdot\bm B=0, \\
\nabla\times\bm E+\frac{\partial\bm B}{\partial t}=0,\\ \nabla\times\bm B=4\pi\bm j+\frac{\partial\bm E}{\partial t}+g_{a\gamma\gamma}\dot{a}(x)\bm B.  \end{array}
\label{eqnew}
\end{equation}


Evidently, Eq. (\ref{eqnew}) does not live place for dual axion electrodynamics and substantially differs from our Eqs. (\ref{eqMlEBw}) and (\ref{eqMEB}). Our approach based on the latter equations \emph{permits} an introduction of effective (fictitious) oscillating magnetic charges and currents in undistorted electromagnetic fields. Equations (\ref{eqMlEBw}) and (\ref{eqMEB}) are self-consistent.  In particular, they correctly describe the emission of axion-distorted electromagnetic waves by oscillating charges, currents, and dipoles, and this description agrees with Eqs.  (\ref{eqtrnsf}) and  (\ref{eqtrnfl}).

We underline that the axion field distorts all electric and magnetic fields but does not change charges and currents. In particular, the axion-photon coupling does not lead to an occurrence of magnetic charges (monopoles) and a shift of positive and negative charges following an appearance of EDMs. The presented approach gives one a \emph{real} physical pictire. However, this approach is equivalent to that consisting in the use of unchanged electric and magnetic fields and the appearance of \emph{effective} magnetic charges \cite{SokolovRingwald} and EDMs \cite{Hill,AxionHillReply,AxionHill2016,AxionCao,AxionZhitnitsky}. The comparison of the two pictures leads to important conclusions. First, extra parameters (magnetic charges and EDMs) can be expressed in terms of electric charges and magnetic dipole moments of particles. Second, the usual quantization rule and other rules considered in Ref. \cite{Heidenreich} do not work in this picture. This property significantly differs our results from those obtained in Refs. \cite{EWitten,Heidenreich}.

In the following, we present quantum-mechanical equations in the Foldy-Wouthuysen representation \cite{JMP,FW} which is an extension of the Schr\"{o}dinger representation to relativistic particles \cite{PRAFW}. The equation of motion for the kinetic momentum follows from Ref. \cite{JMP}:
\begin{equation}   \begin{array}{c}
\frac{d\bm \pi}{dt}=e\bm E'+\frac{e}{4}\Biggl\{\frac{1}{\epsilon},
\Bigl(\bm\pi\times\bm B'-\bm B'\times\bm\pi\Bigr)\Biggr\},
\end{array}  \label{eq35JMP} \end{equation} where $\epsilon=\sqrt{m^2+\bm\pi^2}$. For the electron, $e=-|e|$. We disregard negative total energies. As follows from Refs. \cite{JMP,RPJ}, the corresponding quantum-mechanical equation of spin motion has the form
\begin{equation}
\begin{array}{c}
\frac{d\bm{\Pi}}{dt} =\frac{1}{2}\left\{\left(\frac{\mu_0m}
{\epsilon+m}+\mu'\right)\frac{1}{\epsilon},\Bigl[\bm{\Pi}\times(\bm E'\times\bm \pi-\bm \pi\times\bm
E')\Bigr]\right\}\\+ \left\{\left(
\frac{\mu_0m}{\epsilon}+\mu'\right),[\bm\Sigma\times\bm
B']\right\}\\-  \frac{\mu'}{2}\left\{\frac{1}{\epsilon(\epsilon+m)},\Bigl( [\bm\Sigma\times\bm \pi](\bm \pi\cdot\bm
B')+(\bm B'\cdot\bm \pi) [\bm\Sigma\times\bm \pi]\Bigr)\right\},
\end{array} \label{eq36JMP} \end{equation} where $\mu_0+\mu'=\mu=eg\hbar s/(2mc)$, $\mu_0$ and $\mu'$ are the normal (Dirac) and anomalous magnetic moments, $s$ is the spin number, and $\bm\Pi$ is the polarization operator.

Since $\bm E'$ and $\bm B'$ are defined by Eq. (\ref{eqtrnfl}), Eqs. (\ref{eq35JMP}) and (\ref{eq36JMP}) describe new effects.
Of course, an observer could suppose that the condenser and solenoid are the sources only of electric and magnetic components, respectively, but all spinning particles possess additional axion-induced oscillating EDMs and all charged particles have corresponding magnetic charges. As a result, the dual (inverse) Witten effect takes place. Due to the duality of electrodynamics, this approach brings results equivalent to Eqs.  (\ref{eq35JMP}) and (\ref{eq36JMP}). To solve the problem, we show the classical equations of motion for a particle (dyon) with the electric ($e$) and magnetic ($e^*$) charges and the electric ($d$) and magnetic ($\mu$) dipole moments. The equation for the kinetic momentum reads \cite{Rindler1989}
\begin{equation}
\frac{d\bm\pi}{dt}=e\left(\bm E+\frac{\bm\pi\times\bm B}{\epsilon}\right)+e^*\left(\bm B-\frac{\bm\pi\times\bm E}{\epsilon}\right).
\label{LLL} \end{equation} A comparison with Eqs. (\ref{eqtrnfl}) and (\ref{eq35JMP}) determines the effective (fictitious) magnetic charge:
\begin{equation}
e^*=\frac{g_{a\gamma\gamma}}{2}a(x)e.
\label{fch} \end{equation}
Equations (\ref{eqMlEBw}) and (\ref{fch}) agree.

We can now present the quantity $eA'_\mu$ in dual form. We denote the pseudoscalar field as follows: \begin{equation}A'_\mu=A_\mu+\frac{e^*}{ea(x)}\mathcal{A}_\mu=A_\mu+\frac{g_{a\gamma\gamma}}{2}\mathcal{A}_\mu.\label{ach} \end{equation} In this case, Eq. (\ref{eqnptnt}) is satisfied provided that
\begin{equation}\begin{array}{c}
-\nabla \mathcal{A}_0-\partial_0 \boldsymbol{\mathcal{A}}=\boldsymbol{\mathcal{B}}=a(x)\bm B,\\ \nabla\times\boldsymbol{\mathcal{A}}=-\boldsymbol{\mathcal{E}}=-a(x)\bm E.
\label{fpl} \end{array} \end{equation}
When the derivatives $\nabla a(\bm r,t),~\dot{a}(\bm r,t)$ are negligible, ${\mathcal{A}}_\mu$ is a four-pseudovector.

The classical equation of spin motion obtained in Ref. \cite{EPJPlus} defines the angular velocity of spin precession of a dyon:
\begin{equation} \begin{array} {c}
\bm \Omega=-\frac{e}{m}\left[\left(G+\frac{1}{\gamma}\right){\bm B}-\frac{G\gamma}{\gamma+1}({\bm\beta}\cdot{\bm B}){\bm\beta}\right.\\ \left.-\left(G+\frac{1}{\gamma+1}\right)\bm\beta\times{\bm E}\right]
+\frac{e^*}{m}\left[\left(G^*+\frac{1}{\gamma}\right){\bm E}\right.\\ \left.-\frac{G^*\gamma}{\gamma+1}({\bm\beta}\cdot{\bm E}){\bm\beta}+\left(G^*+\frac{1}{\gamma+1}\right)\bm\beta\times{\bm B}\right],
\end{array} \label{Nelsonf} \end{equation}
where $\bm\beta=\bm\pi/\epsilon$, $\gamma=\epsilon/m$, $G=(g-2)/2$, $g=2mc\mu/(es)$, $G^* =(g^*-2)/2$, and $g^*=-2mcd/(e^*s)$. The equivalence of Eqs. (\ref{eq36JMP}) and (\ref{Nelsonf}) takes place if and only if $G^*=G~(g^*=g)$ and Eq. (\ref{fch}) is satisfied. Evidently, the connection between the effective EDM and the total magnetic moment reads
\begin{equation}
d=\frac{g_{a\gamma\gamma}}{2}a(x)\mu.
\label{fictEDM} \end{equation}

Equations (\ref{fch}) and  (\ref{fictEDM}) describe the dual (inverse) Witten effect. The existence of the original Witten effect \cite{EWitten} can also be confirmed. A motion of the magnetic monopole is given by 
\begin{equation}
\frac{d\bm\pi}{dt}=e^*\left(\bm B'-\frac{\bm\pi\times\bm E'}{\epsilon}\right).
\label{LLm} \end{equation}
The use of Eqs. (\ref{eqtrnfl}) and (\ref{LLL}) results in an appearance of the effective \emph{electric} charge:
\begin{equation}
e=-\frac{g_{a\gamma\gamma}}{2}a(x)e^*.
\label{fWcharg} \end{equation}
The magnetic charge and the EDM of any monopole cause also an appearance of the axion-induced effective magnetic moment.

While the Witten effect is confirmed, we should specify that the effective (fictitious) electric charge is not quantized. We corroborate the statement \cite{AxionCao,AxionZhitnitsky} that the Witten effect leads to an appearance of axion-induced oscillating EDMs of particles with magnetic moments. 

We should note that axion-induced EDMs of nucleons and all other nuclei are also contributed by strong interactions originated from the axion-gluon coupling \cite{GrahamPRDgen,Grahamannurev}:
\begin{equation}
{\cal L}_{g}=\frac{g_{QCD}^2C_g}{32\pi^2f_a}aG_{\mu\nu}\widetilde{G}^{\mu\nu},
\label{eqLgl}
\end{equation} where $g_{QCD}^2/(4\pi)\sim1$ is the coupling constant for the color field, $G_{\mu\nu}$ is the gluon field tensor in QCD, $C_g$ is the model-dependent constant, and $f_a$ is the constant of interaction of axions with matter (axion decay constant). QCD effects are usually much stronger than the corresponding electromagnetic ones. However, their comparison in the considered case significantly depends on the model-dependent parameters. Therefore, axion-induced electromagnetic effects cannot be neglected \emph{a priori} even for strongly interacting particles and nuclei.
We should note that oscillating EDMs caused by the axion-gluon coupling are real (non-fictitious).

The Dirac-Pauli \cite{Pauli} Lagrangian in the presence of the axion field is given by
\begin{equation}\begin{array}{c}
{\cal L}_{DP}=\gamma^\mu\bigl[i\hbar\partial_\mu-eA_\mu-\frac{e^*}{a(x)}\mathcal{A}_\mu\bigr]-m+\frac{\mu'}{2}\sigma^{\mu\nu}F'_{\mu\nu}.
\end{array} \label{electri}
\end{equation}
It can be easily checked following Refs. \cite{RPJ,JMP} that the transition to the Hamiltonian in the Dirac representation and then the relativistic Foldy-Wouthuysen transformation leads to the equations of motion (\ref{eq35JMP}) and (\ref{eq36JMP}).

Due to the duality of electrodynamics, the approach based on introducing effective oscillating magnetic charges and EDMs gives equivalent results. The magnetic charge $e^*$ causes the corresponding effective ``normal'' (``Dirac'') EDM $d_0=e^*\hbar/(2mc)$ for $s=1/2$. The total axion-induced EDM $d=d_0+d'=e^*g\hbar/(4mc)$ is proportional to the \emph{total} magnetic moment $\mu$. One can see that our results have been obtained without addressing to any diagram. The dual approach using effective oscillating magnetic charges and EDMs leads to the following Lagrangian:
\begin{equation}\begin{array}{c}
{\cal L}_{dual}=\gamma^\mu\Bigl(i\hbar\partial_\mu-eA_\mu-\frac{e^*}{a(x)}\mathcal{A}_\mu\Bigr)-m+\frac{\mu'}{2}\sigma^{\mu\nu}F_{\mu\nu}-\frac{d'}{2}\sigma^{\mu\nu}\widetilde{F}_{\mu\nu},\\ \qquad d'=\frac{g_{a\gamma\gamma}}{2}a(x)\mu',
\end{array} \label{elmdual}
\end{equation} where $d'$ is the ``anomalous'' EDM.
The derivation similar to that carried out in Refs. \cite{RPJ,JMP} allows one to obtain Eqs. (\ref{eq35JMP}) and (\ref{eq36JMP}). We underline the full equivalence between the Lagrangians (\ref{electri}) and (\ref{elmdual}). Due to the equivalence of the dual approach to that using distorted fields, experimentalists cannot distinguish between real and effective (fictitious) magnetic charges and EDMs.

Nevertheless, the fact that the magnetic charges and EDMs are effective (fictitious) but not real is crucial. Just this fact allows us to avoid known difficulties with a coexistence of electric and magnetic charges. The controversiality of dual axion electrodynamics has been recently shown in Ref. \cite{Heidenreich}. In reality, electric and magnetic fields \emph{distorted} by axions interact with electric charges, magnetic moments and, for nuclei, also with oscillating EDMs caused by axion-gluon coupling. One can also use the equivalent picture based on \emph{undistorted} electric and magnetic fields and the appearance of effective (fictitious) oscillating magnetic charges and EDMs. However, in this picture neither the usual quantization rule nor other rules (see Ref. \cite{Heidenreich}) work. Due to this circumstance, an introduction of effective magnetic charges and EDMs becomes noncontradictory.

For leptons, one previously studied the axion wind effect (see Refs. \cite{PhysUsp,axiondJEDI,PospelovAxion,DereviankoAxion,StadnikFlambaumAxion,axionewind,BudkerRMP,FrozEDMPRD,EPJC2022}). This effect appears when axions move relative to a particle. The corresponding Lagrangian density is given by
\begin{equation}\begin{array}{c}
{\cal L}_{wind}=g_{aNN}\gamma^\mu\gamma^5\Lambda_\mu,
\\ \Lambda_\mu=(\Lambda_0,\bm\Lambda)=\partial_\mu a=(\dot{a},\nabla a),
\end{array} \label{electric}
\end{equation}  where 
$g_{aNN}$ is the model-dependent constant defining axion-particle coupling. ${\cal L}_{wind}$ contains the four-pseudopotential $\mathfrak{A}_\mu$ of axion field. It is pseudoscalar but not pseudovector and should satisfy the condition $\partial_\mu\mathfrak{A}_\nu-\partial_\nu\mathfrak{A}_\mu=0$. This condition is valid and $\mathfrak{A}_\mu=g_{aNN}\gamma^5\partial_\mu a$. For leptons, the list of axion-induced effects containing effective (fictitious) time-dependent magnetic charges and EDMs and the axion wind effect is \emph{exhaustive}.

The dual (inverse) Witten effect defining effective magnetic charges of electrically charged particles opens new exciting possibilities to search for dark matter axions and ALPs. Importantly, corresponding experiments can be non-resonant contrary to monitoring the spin motion. As a result, they are exceptionally important for a search for axions and ALPs with very small masses and, therefore, oscillation frequencies $\omega_a=E_a/\hbar\approx m_ac^2/\hbar$. We can propose to study a passage of strongly decelerated electrons or positrons through a solenoid. In this case, the oscillating electric field $\bm E'$ changing a particle velocity appears [see Eq. (\ref{eqtrnfl})]. When $m_a$ is very small, this field is almost unchanged during the particle passage through a solenoid and conditions a particle acceleration. Equations (\ref{axion}) and (\ref{eqtrnfl}) show that this acceleration depends on a moment of measurement and can be detected. The effect does not need any resonance fields and is detectable on condition that $1/T\gtrsim\omega_a$, where $T$ is the time of the particle passage through the solenoid. When this inequality is not satisfied, one can either use faster particles or fulfill experiments already started by CASPEr \cite{PhysRevXCASPEr,PhysRevLettCASPEr} and JEDI \cite{axiondJEDI} collaborations and other experimental groups \cite{Grahamannurev,BudkerRMP,AxionEDMneutronPXR,AntiprotonAxion}. In the latter case, taking into account the considered effects is very important, especially for experiments with atoms and molecules. In addition,
bounds obtained in experiments with haloscopes can be easily joined with bounds found in such experiments.

In summary, we have developed the approach which is based on the covariance of Lagrangians and describes a distortion of \emph{any} electromagnetic field by a pseudoscalar axion one. As a result, electric and magnetic fields acquire oscillating magnetic and electric components, respectively. We have substantiated the need for correction of the generally accepted total Lagrangian density. The Maxwell-like equations have been derived. One can also use the equivalent approach based on introducing effective (fictitious) oscillating magnetic charges and EDMs in undistorted electromagnetic fields. We have rigorously determined the relations between these magnetic charges and EDMs, real electric charges and magnetic moments, and the axion-photon coupling constant. The results obtained open a possibility to compare a sensitivity of search for dark matter axions (and APLs) in optical experiments and experiments with massive particles. In addition, the dual (inverse) Witten effect defining effective (fictitious) magnetic charges of electrically charged particles opens new exciting possibilities to search for axions and ALPs with very small masses in nonresonance experiments. We propose to detect an appearance of an oscillating electric field changing a particle velocity during a passage of strongly decelerated electrons or positrons through a solenoid.

\section*{Data availability statement}

All data that support the findings of this study are included within the article.


\begin{acknowledgments}
This work was supported by the Russian Science Foundation (Grant No. 25-72-30005). The author is grateful to D. Budker for paying his attention to Refs. \cite{Hill,FlambaumComment} and C. T. Hill and N. N. Nikolaev for valuable comments and discussions.\end{acknowledgments}


\begin{thebibliography}{}
\bibitem{footnote}
The term ``fictitious'' does not mean ``nonexistent''. For example, a centrifugal force acting in a rotating frame is fictitious while it is perfectly felt by a rotating observer.
\end{thebibliography}
\end{document}